\documentclass{article}
\usepackage[paperwidth=235.9mm, paperheight=279.4mm, margin=50mm]{geometry}
\usepackage{kr}

\usepackage{times}
\usepackage{soul}
\usepackage{url}
\usepackage[hidelinks]{hyperref}
\usepackage[utf8]{inputenc}
\usepackage[small]{caption}
\usepackage{graphicx}
\usepackage{amsmath}
\usepackage{amsthm}
\usepackage{amssymb}
\usepackage{booktabs}
\usepackage{algorithm}
\usepackage{algorithmic}
\usepackage{xspace}
\usepackage{booktabs}
\usepackage{decision-table}
\usepackage{enumitem}

\usepackage{idpz3syntax}
\usepackage{customcommands}
\usepackage{tablefootnote} 
\usepackage{wrapfig}

\newtheorem{example}{Example}

\usepackage{todonotes}
\newcommand{\marc}[1]{{ \color{red} (MD) #1 } }

\newcommand{\fodot}{FO($\cdot$)\xspace}
\newcommand{\fodotp}{FO[$\cdot$]\xspace}

\title{Interactive configurator with \fodot and IDP-Z3}

\author{%
Pierre Carbonnelle$^{1,3}$\and
Simon Vandevelde$^{2,3,4}$\and
Joost Vennekens$^{2,3,4}$\and
Marc Denecker$^{1,3}$ \\
\affiliations
$^1$KU Leuven, Dept. of Computer Science, Celestijnenlaan 200, 3001 Heverlee, Belgium\\
$^2$KU Leuven, De Nayer Campus, Dept. of Computer Science,\\ J.-P.-De Nayerlaan 5, 2860 Sint-Katelijne-Waver, Belgium\\
$^3$Leuven.AI -- KU Leuven Institute for AI, B-3000, Leuven, Belgium \\
$^4$Flanders Make@KU Leuven\\
\emails
\{pierre.carbonnelle, s.vandevelde, joost.vennekens, marc.denecker\}@kuleuven.be
}

\begin{document}

\maketitle

\ignore{
    \fodot (aka FO-dot) is a model-based Knowledge Representation language that extends classical first-order (FO) logic. 
    As in FO, formulas in \fodot express knowledge about possible states of affairs; formally, they define a set of \emph{models}, i.e., of possible worlds that satisfy the formula.
    Knowledge bases in \fodot cannot be run, but they can serve as input to generic reasoning methods. 
}

\begin{abstract}
Industry abounds with interactive configuration problems, i.e., constraint solving problems interactively solved by persons with the assistance of a computer.
The computer program, called a configurator, needs to perform a variety of reasoning tasks with the (often incomplete) information that the user provides.
Imperative programming approaches make such systems difficult to implement and maintain.
Knowledge-based configurators have been proposed to help engineers solve such problems, but many challenges remain.

We present IDP-Z3, a new reasoning engine for the \fodot KR language, and we report on its use for building configurators automatically from a knowledge base.
\end{abstract}

\ignore{  ** Old Abstract
    \fodot (aka FO-dot) is a language that extends classical first-order logic with constructs to allow complex knowledge to be represented in a natural and elaboration-tolerant way. 
    
    IDP-Z3 is a new reasoning engine for the \fodot language: it can perform a variety of generic computational tasks using knowledge represented in \fodot.
    It supersedes IDP3, its predecessor, with new capabilities such as support for linear arithmetic over reals and quantification over concepts.
    
    We present four knowledge-intensive industrial use cases, and show that IDP-Z3 delivers real value to its users at low development costs: it supports interactive applications in a variety of problem domains, with a  response time typically below 3 seconds.
    
}

\ignore{
    Paper on the IDP-Z3 ``ecosystem'' (IDP-Z3, IC, DMN, CNL, ...)
    
    First: focus on \fodot and IDP-Z3 (= the core technology)
    \begin{itemize}
        \item Open-ended and extensible
        \item On the modelling side, allows us to use DMN, CNL, ... 
        \begin{itemize}
            \item Thanks to modular format of \fodot and its basis in classical logic
        \end{itemize}
        \item On the interaction side, allows us to
        \begin{itemize}
            \item offer inferences independent of knowledge
            \item besides pre-built inferences, more can be added for specific cases(cfr Bart Coppens -- Appendix A)
            \item realise other usages for our knowledge (e.g. DMN-IDP and DMN verification) instead of tunnel-visioning on one specific usage
        \end{itemize}
    \end{itemize}
    
    IDP-Z3 combines rule-based and constraint-based logic
    
    Key: open-mindedness, both on ``how to construct knowledge'' and ``what to do with it''.
    Both creation of KB and interaction with it (IC) is user-friendly.
    But can also be embedded in concrete tools (e.g., Bart Coppens)
    
    History of \fodot, IDP (with awards).
    \cite{decatPredicateLogicModeling2018}
    
    Also mention Knowledge Base Paradigm \cite{DBLP:conf/iclp/DeneckerV08}
    
    Mention ``old'' IC (i.e., AutoConfig) here as well?
}

\section{Introduction}

An interactive configuration problem ~\cite{DBLP:journals/ai/McDermott82,DBLP:conf/aaai/MittalF90,DBLP:journals/expert/FleischanderlFHSS98,DBLP:conf/aaai/Junker04,DBLP:conf/cp/HadzicA05} is an interactive version of a constraint solving problem. 
With the assistance of a computer, one or more users search for a  configuration of objects and relations between them that satisfies a set of constraints. Industry abounds with interactive configuration problems: configuring composite physical systems such as cars and computers, or composite non-physical constructs such as insurance contracts, loans and schedules.

An interactive configuration system, aka a \emph{configurator}, needs to provide many functionalities to its user: 
asking the user relevant questions,
computing the consequences of choices made by the user,
providing explanation for these consequences,
handling missing information, being flexible on data entry (i.e., not imposing a strict order on data entry), 
backtracking if the user regrets some choices,
completing a partial configuration to minimize a cost function,
checking the validity of a fully specified configuration, etc.


Building configurators with all these functionalities is renowned for being difficult, and no method to build them is broadly accepted~\cite{Felfernig14,DBLP:journals/jlp/AxlingH96}. 
%
Building configurators using standard imperative programming is often a nightmare~\cite{DBLP:journals/cacm/BarkerO89,DBLP:conf/hicss/PillerHIS14} because (1) many functionalities need to be provided, (2) they are complex to implement, and (3) the resulting programs turn out to be extremely difficult to maintain when existing constraints of the application domain change or   new  constraints emerge. 
Indeed, the change of one constraint may require changes in many different (and often non-obviously related) parts of the program code.

For this type of applications, it is  natural to look for {\em declarative solutions}. Rephrasing \citeauthor{felfernig2014configuration} (\citeyear{felfernig2014configuration},~p192), the research question is how can we build interactive configuration systems in which the knowledge of the application domain (the constraints) can be expressed in a formal modelling language, such that the system can provide the aforementioned functionalities by performing the required form of reasoning on the knowledge base?

\ignore{Solving these challenges would allow the creation of machines at the ``fourth level'' of the scale presented by \citeauthor{mccarthy1989artificial} (\citeyear{mccarthy1989artificial}).
    Machines at the fourth level are the most intelligent.  
    The fourth level ``involves representing general facts about the world as logical sentences. [..] The facts would have the neutrality of purpose characteristic of much human information.'' 
}


These challenges make interactive configuration an ideal application domain for the {\em  Knowledge Base Paradigm } \cite{DBLP:conf/iclp/DeneckerV08}. This paradigm returns to the classical view of logic as a formal language to express {\em pure knowledge} in a declarative way. 
In this paradigm, an application is developed by writing a  {\em  declarative knowledge base}, and by using a knowledge base system that can perform a variety of tasks  using the same declarative knowledge base by applying various forms of {\em inference}. Phrased in the terminology of \citeauthor{mccarthy1989artificial} (\citeyear{mccarthy1989artificial}) 
the knowledge in the knowledge base has a {\em neutrality of purpose, like any human knowledge}, and the knowledge base system is at McCarthy's {\em fourth level of the scale of intelligent logic machine}. Machines at the fourth level go beyond deductive reasoning and theorem proving and support various forms of reasoning and inference for a variety of purposes.

The challenges of building such solutions for interactive configuration  are high. First, some of the domain knowledge may be complex. Many interactive configuration problems include complex  constraints: various sorts of quantification, aggregates, definitions (sometimes inductive), etc. Thus, the declarative language must be expressive; more expressive than FO predicate logic. Second,  many functionalities must be provided in interactive configuration, and must be implemented using various forms of inference on the given knowledge base. This is complex.  



The KB paradigm was followed by~\citeauthor{VanHertum2017}~(\citeyear{VanHertum2017}) in a prototype configurator called AutoConfig. 
This configurator allowed the user to enter data in any order, and, after any data entry, performed a variety of computations to assist the user in his exploration of the solution space. 
To represent complex knowledge, it used the \fodot language~\cite{tocl/DeneckerT08}, an extension of First Order logic with inductive definitions and other language constructs to make it more expressive.
To perform the computations, it used the IDP3 reasoning engine for \fodot~\cite{WarrenBook/DeCatBBD18}.
The user interface was generated automatically from the knowledge base: no programming was required, significantly reducing development costs~\cite{Deryck2019}. 

In this paper for the Applications and Systems track, we describe the successor of the IDP3 reasoning engine, called IDP-Z3, and its use in interactive configuration.

\ignore{
    IDP-Z3 improves upon IDP3 on a range of different aspects, such as more systematic syntax, better support for identifiers and constructors, a distinction between environmental versus decision facts, etc.  
    But, the main feature where IDP-Z3 makes the difference with the finite domain reasoner IDP3 in real applications, is its arithmetic reasoning skills, which are borrowed from the underlying  SMT solver Z3. Arithmetic is important in many real applications, and the improved reasoning skills of the new system often makes the difference between being applicable or not. 
}
IDP-Z3 improves on IDP3 in significant ways.
First, IDP3 could only reason on finite domains, severely limiting its arithmetic capabilities.  By contrast, IDP-Z3 supports non-linear arithmetic over the rationals, opening up its field of application.
Second, the configurator allows for a safer exploration of the solution space by the user, by distinguishing variables that the user controls (called \emph{decision} variables) from those that he does not (called \emph{environmental} variables), and by distinguishing theories about them.\footnote{We have recently submitted a paper to ICLP that presents the theory behind the distinction between environmental and decision variables in detail.  The focus of that paper differs significantly from the focus of this paper, which is for the Applications and Systems track.} 
IDP-Z3 also makes incremental improvements to the \fodot language, such as more systematic syntax and better support for identifiers and constructors.

We begin by presenting \fodot in Section~\ref{sec:fodot}.
Next, we present the IDP-Z3 engine and its features in Section~\ref{IDP-Z3}, followed by Section~\ref{sec:IC} in which we expand on the Interactive Consultant, its generic configurator interface.
As an empirical evaluation of the system, we present industrial use cases in Section~\ref{sec:eval}.
Finally, we compare IDP-Z3 to other reasoning engines for model-based KR languages in Section~\ref{sec:related}, and conclude in Section~\ref{sec:conclusion}.

In short, the contributions of this paper are:
\begin{itemize}
    \item an overview of the \fodot Knowledge Representation language;
    \item the presentation of IDP-Z3 and the Interactive Consultant;
    \item an evaluation of IDP-Z3 in industrial use cases that demonstrate the benefits of creating interactive configurators using IDP-Z3;
    \item a qualitative comparison between IDP-Z3 and other reasoning engines.
\end{itemize}

\ignore{ 
    \section{Introduction}
    
    \citeauthor{mccarthy1989artificial} (\citeyear{mccarthy1989artificial}) presents four possible levels of use of logic in Artificial Intelligence.
    Intelligent machines at the first level, such as neural networks, do not use logic sentences at all. 
    At the second level, machines use logic sentences to represent facts from which they reach conclusions using ad-hoc procedures, typically written in imperative programming languages, without the generality of ordinary logical inference.
    
    Machines at the third level use logical deduction to reach conclusions.  
    He cites Prolog as one of the languages used to program them.  
    Such machines are rather specialized: ``the facts of one program usually cannot be used in a database for other programs.''
    This is a result of their fixed deduction strategy: 
    because Algorithm = Logic + Control~\cite{kowalski1979algorithm}, one has to use a new set of logic rules to create an algorithm for a new task in the same problem domain.
    
    By contrast, machines of the fourth and most advanced level do not have a fixed deduction strategy.  
    The fourth level ``involves representing general facts about the world as logical sentences. [..] The facts would have the neutrality of purpose characteristic of much human information. [..] A key problem for achieving the fourth level is to develop a language for a general, common-sense database.''.
    
    The IDP-Z3 system seeks to address that challenge: it is designed so that a) one can express knowledge about possible worlds using logical sentences, and b) this knowledge can be used for many different computational tasks.
    To distinguish it from inference engines at the third level, we call it a ``reasoning engine''.
    Reasoning engines enable the Knowledge Base paradigm~\cite{DBLP:conf/iclp/DeneckerV08}, in which systems store declarative domain knowledge, and use it to solve a variety of problems.
    This approach can significantly reduce the development and maintenance costs of intelligent machines~\cite{Deryck2019}.
    
    \ignore{
        While the semantics of third-level formalisms is proof-based (in the sense that the meaning of a program is a set of inference rules used to build a proof using a fixed control strategy), the semantics of \fodot is  model-based, in the sense that it defines the satisfaction relation between a formula and the set of its \emph{models}, i.e., the set of possible worlds satisfying it.

        IDP-Z3 uses the \fodot language to express knowledge.
        SMT-LIB~\cite{SMT} and ASP~\cite{ASP} are other languages that can be used to develop 4th-level machines.
        Efficient reasoning with them is made possible by recent advances in SAT, SMT and ASP solvers.
    } 
    
    In this ``System Description'' paper, we present IDP-Z3 and various tools and extensions built around it.
    In particular, we demonstrate that IDP-Z3 allows users to leverage their domain knowledge to produce flexible interactive systems that offer all the necessary functionality and computational efficiency to handle real-world problems. 
    There are several aspects to this:
    
    \begin{itemize}
        \item The \fodot language is important to allow complex knowledge to be represented in a natural and elaboration-tolerant knowledge base. 
        \item The modular and classical nature of \fodot make it easy to extend a KB with parts that are written in more user-friendly notations such as DMN or Controlled Natural Language.
        \item The range and speed of the generic reasoning algorithms offered by IDP-Z3 suffices to implement a large class of interactive applications in an efficient way.
    \end{itemize}
    
    We present four knowledge-intensive use cases as proof for our claims: they show that (1) real users are indeed able to participate in the construction of the KB, (2) that IDP-Z3 delivers significant value to these users, at low development costs, (3) that the IDP-Z3 system, while not the most efficient solver that exists, is able to handle the real-world instances that the users want to tackle.
    
    We begin by elaborating on \fodot and alternative formalisms in Section~\ref{sec:fodot}.
    Next, we present the IDP-Z3 engine and its features in Section~\ref{IDP-Z3}, followed by Section~\ref{sec:IC} in which we expand on the Interactive Consultant, a generic, user-friendly interface to solve real-world problem using the reasoning power of IDP-Z3.
    As an empirical evaluation of the system, we report on four knowledge-intensive industrial use cases in Section~??, and demonstrate the benefits of creating interactive applications using IDP-Z3.
    \todo{update with new section}
    Finally, we compare IDP-Z3 to other reasoning engines for model-based KR languages in Section~\ref{sec:eval}, and conclude in Section~\ref{sec:conclusion}.
    
    In short, the contributions of this paper are:
    \begin{itemize}
        \item an overview of the \fodot Knowledge Representation language;
        \item the presentation of IDP-Z3 and the Interactive Consultant;
        \item a summary of case studies involving IDP-Z3, to support our claims;
        \item a qualitative comparison between IDP-Z3 and other reasoning engines.
    \end{itemize}
}

\section{\fodot}
\label{sec:fodot}

\fodot (aka FO-dot) \cite{DeneckerCL2000,tocl/DeneckerT08} is the language used to represent knowledge in the IDP-Z3 reasoning engine.   

\fodot is based on first order predicate logic (classical FO).
Classical FO has a simple and natural informal semantics,  and a clear and mathematically precise formal semantics.
The semantics of the logic connectives $\land, \lor, \neg, \Rightarrow, \Leftrightarrow, \forall, \exists$ are clear, precise, simple and natural.
The formal semantics of FO defines the set of structures satisfying a formula, i.e., it defines the \emph{satisfaction} relation.
The informal semantics of FO defines the \emph{abstraction} relation, i.e., the relation between components of the formal semantics and the components in the problem domain that they abstract.
A structure satisfying a formula, i.e., a model, is the abstraction of a possible state of affairs.

Classical FO is used successfully for the mathematical modelling of science.  
The use of its formal and informal semantics is essential to its success.
A theory, i.e., a set of FO formulas, represents knowledge about a problem domain by partitioning the set of states of affairs in two sets: the ones whose abstraction satisfies the theory, and the ones whose abstraction does not.
The more knowledge a theory contains, the fewer models it has, and the fewer states of affairs it deems possible.
Modelling of a problem domain is performed by adding knowledge to the theory, i.e., by reducing the number of its models.
Modelling is complete when the set of models is precisely the set of abstractions of the possible states of affairs.

Many reasoning methods and tools exist for FO.
For example, deductive methods have been developed to find formulas entailed by a theory, or to prove that two theories are equivalent. 
These concepts of entailment and equivalence are defined in terms of the distinction between models and non-models.
But the knowledge in a FO theory can be used to solve many other computational problems beyond computing entailment and equivalence, as explained in the introduction. 
We shall describe many such inferences in the next section.

Despite its success in formulating science, FO has many weaknesses for representing human expert knowledge in complex, heterogeneous real-world applications. 
For this reason, \fodot extends FO with a range of language constructs such as types, identifiers, binary quantification, arithmetic, aggregates, and definitions while keeping the essential properties of classical FO listed above. 
\fodot defines the semantics of these constructs by extending the definition of the classical satisfaction and abstraction relations.

The following example illustrates the use of types, arithmetic, aggregates and definitions in an \fodot theory.

\newcommand{\Company}{\m{\mathit{Company}}}
\newcommand{\Cont}{\m{\mathit{Cont}}}
\newcommand{\Real}{\m{\mathit{Real}}}
\newcommand{\OwnsSh}{\m{\mathit{OwnsSh}}}
\newcommand{\American}{\m{\mathit{American}}}
\begin{example}[Company control] \label{ex:company:control}
Consider the informal statement ``every American company is controlled by other American companies'', 
where ``control'' is defined as follows:
``A company A controls company B if the total sum of the shares in company B owned by A or by companies controlled by A is more than 50\%.''
 
We use the following vocabulary:
\begin{itemize}
    \item \Company: a type, i.e., a subset of the domain of discourse;
    \item ${Cont(\Company, \Company)}$: a predicate symbol; $\Cont(x,y)$ is informally interpreted as ``company {x} controls company {y}'';
    \item ${\OwnsSh: \Company \times \Company \rightarrow \Real}$:  a function symbol: $\OwnsSh(x,y)=s$ is informally interpreted as ``company $x$ owns s \% of the shares of company {y}'';
    \item $\American(\Company)$: $\American(x)$ is informally interpreted as ``company {x} is American''.
\end{itemize}

The first \fodot formula below defines $\Cont$ and the second formulates the informal statement:
\begin{align*}
    \forall a \forall b &(\Cont(a,b) \leftarrow  50 < \SUM{c}{c=a \lor \Cont(a,c)}{\OwnsSh(c,b)}	) \\
    \forall c, c1 &(\Cont(c,c1) \land \American(c1) \Rightarrow \American(c))
\end{align*}
\end{example}

\fodot is unique as a KR language for its attempt to combine the strengths of FO and of logic programming.  
FO cannot represent inductive definitions, such as the definition of $\Cont$ in our example.
By contrast, logic programming can represent inductive definitions, but has difficulty representing other forms of knowledge, such as the statement that every American company must be controlled by American companies.

We highlight the main features of \fodot below.    
The syntax of the concrete logic used  in IDP-Z3, called \fodotp, is documented online\footnote{\url{http://docs.idp-z3.be/en/latest/FO-dot.html}}.

\paragraph{Types and quantification}
In real applications, the universe tends to be  heterogeneous, and there are virtually no properties or concepts shared by all objects in the domain. In our mind, the universe is  partitioned into different more homogeneous {\em categories} of objects, each with its own class of attributes and relations.  A type system serves to support this view,  leading to more natural and precise KR.
(Types are also called ``sorts'' in the literature~\cite{wang1952logic})

Besides boolean, integer and real built-in types, \fodotp allows the creation of custom types, e.g., \idp{Person}.
Predicates and functions are declared in the vocabulary, with a \emph{type signature}, e.g., \idp{colorOf: Node -> Color}.
In the quantified formula \idp{!x \\in T: p(x)}, \idp{x} ranges over the extension of type \idp{T}.
Types are used to syntactically verify that formulae are \emph{well-typed}, helping detect common errors.

It is also convenient to quantify over the set interpreting a predicate.
This formula uses \emph{binary quantification} to quantify over the edges: \idp{!(n1,n2) \\in edge: p(n1,n2)}.

\ignore{
    \paragraph{Axioms} \marc{USE THE EXISTING TERMINOLOGY. WHEN YOUR WRITE A PAPER, GO BACK TO LOOK UP TERMINOLOGY OF A RECENT PAPER, OR HERE THE COURSE MODELLING OF COMPLEX SYSTEMS, AND USE THE TERMINOLOGY THAT EXISTS. THIS IS IMPORTANT FOR CONTINUITY AND UNDERSTANDING, AND FOR THE QUALITY OF SCIENTIFIC TEXT. DONT IMPROVISE A NEW TERMINOLOGY. Here, new terminology is invented:"axioms, assertions". In past papers, these terms were not used, but they are introduced here. THis is not a good idea. In a science, one creates a terminology that is to be precisely understood. It is therefore not wise to improvise other terms to talk about the same concepts. Please adhere to the existing terminology. 
    }
    
    The first class of assertions in an \fodot theory is the class of axioms.
    Axioms are logic sentences that are true in any possible state of affairs. \marc{Mixing what an axiom  with its purpose. ``Axiom'' is not the right terminology. 
    What need to be said here in my opinion, is that FO(.) theories consist of FO(.) formulas and FO(.) definitions. The FO(.) formulas are FO formulas extended with other language constructs such as typing, binary quantification, aggregates, as illustrated below.    }
    
    In the graph coloring example, the axiom stating that adjacent nodes have different colors is expressed as:
    \begin{lstlisting}[language=IDP]
    !(n1, n2) in edge: colorOf(n1)~=colorOf(n2).
    \end{lstlisting}
    
    \ignore{
        The voting law above is an example of an axiom.
        A voting law that does not make voting mandatory can be expressed in another axiom using material implication:
        \begin{lstlisting}[language=IDP]
            vote() => 18=<age().
        \end{lstlisting}
        (Notice that the voting obligations and permissions are represented without any modal operator, unlike formulations in deontic logic~\cite{von1951deontic}.)
    }
    
    \marc{As for how to call the elements of an FO(.) theory. You proposed ``assertion'' and now ``axiom''. I suggest to use ``formulas'' and ``definitions''.  Notice that in course KR, distinction is made between assertive and definitional K; so definitions should not be assertions. } 
}

\paragraph{Definitions (inductive or non-inductive)}

Besides logic formulas, an \fodotp knowledge base may include (possibly inductive) definitions of sets and relations.
Definitions are an important, common and precise form of human knowledge.
Definitions define one or more  defined concepts in terms of other {\em parameter} concepts, meaning that they specify a unique value for  the defined symbol(s), given a value for  of its parameter concepts. 

It is well known that FO cannot represent inductive definitions, such as the transitive closure of a graph.
By contrast, \fodotp can represent such definitions using an extension of FO for inductive definitions called FO(ID)
~\cite{DeneckerCL2000}.
Models of the definition are the structures in which the interpretations of the defined symbols are identical to the results of an iterative construction process that uses the interpretation of the parameters of the definition in the structure.
\ignore{
    However, consistent with the model-based approach, such definitions allow reasoning with any partial knowledge, in any direction.
    For example, it allows finding all the graphs that have a given transitive closure.
}

Definitions are often formulated in natural languages as a set of ``rules'' specifying necessary and sufficient conditions for the definiendum to hold.
\fodotp definitions are also of this form, as illustrated in Listing~\ref{lst:1}.
The definiens, i.e., the body of a rule, can be any FO formula.
The formalism of definitions in \fodotp is elaboration tolerant in the sense that one can easily add a rule to a definition.
\begin{lstlisting}[language=IDP, label={lst:1}, caption=Multi-rule definition]
{ ! n1,n2: reachable(n1,n2) <- edge(n1,n2).
  ! n1,n2: reachable(n1,n2) <- 
      ? n3: edge(n1,n3) & reachable(n3,n2).
}
\end{lstlisting}

By default, the formal semantics of definitions is the well-founded semantics, but other semantics can be used (namely the completion, co-inductive and Kripke-Kleene semantics).
Because rules are part of definitions in \fodotp, the head of a rule must be a single atom (in contrast to ASP which allows a disjunction in the head of so-called choice rules).
Unlike in default logic~\cite{DBLP:journals/ai/Reiter80}, exceptions to a rule must be explicitly stated in the rule (possibly in the form of a predicate defined separately).
Unlike in defeasible logic~\cite{DBLP:conf/inap/Nute01}, rules do not have any priority ordering.

\paragraph{Data theory}
FO is not well suited to express simple data about a concrete state of affairs. 
Unique Names (UNA), Domain Closure (DCA), and Completion (CA) Axioms are needed, but stating them explicitly is cumbersome.
For example, the UNA consists of a number of axioms quadratic in the number of constant symbols.

To address this issue, \fodotp theories may use enumerations.
Enumerations define the set of identifiers interpreting a type (under UNA and DCA), and the sets and functions interpreting predicate and function symbols (under DCA and CA)\footnote{A data theory defines a (partial) structure.  For historical reasons, we call it a \lstinline|structure| in \fodotp.}.
For example, \idp{Node:=\{A,B\}} and \idp{colorOf:=\{A->Red, B->Blue\}} define the interpretation of type \idp{Node} and of function \idp{colorOf}.

Identifiers differ from constants in that they are subject to UNA.  To make that distinction explicit, \fodotp now requires parenthesis after constants (e.g., \idp{weight()}), but not after identifiers.

\paragraph{Arithmetic}
\fodotp supports formulae with the 4 arithmetic operations on integers and reals (addition, subtraction, multiplication and division), as well as the comparison operators (equality, disequality and inequalities).
It does not support transcendental functions (\idp{log,exp,sin,cos,..}).

\paragraph{Cardinality and Aggregates}
The number of elements of finite type T that satisfy a property \idp{p} is formulated in \fodotp as:
\begin{lstlisting}[language=IDP]
    #{el \in T: p(t)}
\end{lstlisting}

Similarly, \idp{sum(lambda x \\in T: f(x))} denotes the sum of \idp{f(x)} for each \idp{x} in \idp{T}.
The minimum and maximum aggregates are denoted \idp{min}, \idp{max}.
Cardinality and aggregate expressions can occur in formulas and in the body of rules.

\paragraph{\lstinline|Concept| as a type}
Sometimes, it is necessary to reason about the concepts in an ontology. 
Informally, the concept behind an ontology symbol is its informal meaning.
\fodotp introduces the type \lstinline|Concept| (whose extension is the set of concepts in the ontology of the problem domain), and the ``\lstinline|$|'' operator that maps a concept to its interpretation. 

For example, one might want to count the number of node attributes that a node \idp{n} has, such as being \idp{big}, \idp{bright} or \idp{heavy}.
Such attributes are normally represented by predicates, e.g., \idp{big:Node->Bool}.
The number of node attributes for a node \idp{n} can be formulated as:
\begin{lstlisting}[language=IDP]
    #{a \in node_attribute: $(a)(n)}
\end{lstlisting}
where \idp{node_attribute:=\{`big, `bright, `heavy\}} is the set of node attributes.

This concise formalism avoids the need to reify the node property predicates, i.e., defining a \idp{type Property:=\{Big,Bright, Heavy\}}, and a predicate \idp{has:Node*Property->Bool}.
Reification is not elaboration tolerant~\cite{mccarthy1998elaboration} because it requires rewriting every formula involving the node properties.

\section{IDP-Z3}
\label{IDP-Z3}

IDP-Z3 is a reasoning engine that can perform a variety of reasoning tasks on knowledge bases in the \fodot language.
It is designed according to the knowledge base paradigm: knowledge of a domain is expressed in an expressive formal language, \fodot, and used by generic algorithms to solve a variety of problems in the problem domain.
The re-use of these algorithms across problem domains significantly reduces system development and maintenance costs.

\ignore{
    FOLASP~\cite{VanDessel2021} is another reasoning engine for \fodot.  
    It uses ASP-Core-2 solvers as back-end.
    Its reasoning capabilities are limited to one reasoning task, however: model expansion.
}


The following generic forms of reasoning, i.e., inferences, are supported by IDP-Z3:

\begin{itemize}
    \item \textbf{Model checking}
        Takes a theory T and a total structure S, and verifies that S is a model of T.
    \item \textbf{Model expansion}
        Takes a theory T and a partial structure S, and computes a model S' of T that is more precise\footnote{I.e., every expression having a definite value  in S has the same value in S'} than S if one exists.
    \item \textbf{Possible values}
        Takes a theory T, a partial structure S, and a term t, and computes the list of values of t in the model expansions of T and S.
    \item \textbf{Propagation}
        Takes a theory T and a partial structure S, and computes, for every tuple in the domain of each symbol, whether the value of the tuple applied to the symbol is the same in every model expansion of T and S.
        This computation is also called ``backbone computation'' in the literature (e.g., \cite{zhang2019smtbcf}).
    \item \textbf{Explanation}
        Takes a theory T, a partial structure S and a literal L obtained by propagation, and computes an explanation for L in the form of a minimal set of elements in $T \cup S \cup \{\lnot L\}$ that is inconsistent.
        To perform this inference, we use an ``unsat\_core'' algorithm~\cite{biere2009handbook}.
        \footnote{To create step-by-step explanations, \citeauthor{DBLP:conf/ecai/step-wise} (\citeyear{DBLP:conf/ecai/step-wise}) used IDP3 to find step-wise explanations.
        This method has not yet been ported to IDP-Z3.}
    \item \textbf{Optimization}
        Takes a theory T, a partial structure S and a term, and computes the minimal value of the term in the set of all model expansions of T and S.
    \item \textbf{Relevance}
        Takes a theory T and a partial structure S, and determines the atoms that are irrelevant (or ``do-not-care'').
        This inference is described further below.
\end{itemize}

\ignore{\paragraph{Abstract model Generation}

    \paragraph{Other reasoning tasks}
    While not natively supported in IDP-Z3, other reasoning tasks can be developed around IDP-Z3.
    For example, one could compare two \fodot formulations, and show models where they differ, as in the Intelli-Select use case described in Section~\ref{sec:intelli-select}.
    One could also verify the completeness of definitions (or of DMN tables), or generate test cases for a KB.
}

IDP-Z3 can be run at the command line, or integrated in a Python application as a Python package downloadable from the Python Package repository\footnote{\url{https://pypi.org/project/idp-engine/}}.
Computations can also be run online via a \nobreak{webIDE}\footnote{\url{https://interactive-consultant.idp-z3.be/IDE}}.
It is open source\footnote{\url{https://gitlab.com/krr/IDP-Z3}} under the LGPL 3 license.

IDP-Z3 is the successor of IDP3~\cite{de2013idp3}, which used a custom SAT solver, called minisat(ID)~\cite{de2014minisat}.
Hence, its support for arithmetic was limited.
By contrast, IDP-Z3 uses an off-the-shelf SMT solver, Z3~\cite{de2008z3}, which supports reasoning over linear arithmetic.  

A challenge in migrating IDP3 to the Z3 SMT solver was in re-implementing some specific capabilities of minisat(ID).
In particular, minisat(ID) used custom procedures to handle (inductive) definitions and to compute relevance~\cite{ijcai/JanotaM15}.
We now describe how these capabilities are implemented in IDP-Z3.

\textbf{Non-inductive definitions} are translated to formulae acceptable by Z3 using Clark's completion semantics~\cite{DBLP:conf/adbt/Clark77}: a defined atom must have the same truth value as the disjunction of the body of the rules defining it.
For example, suppose that the graph is directed.  
Whether two nodes are connected is defined as follows:

\begin{lstlisting}[language=IDP, label={lst:2}, caption=Definition]
  { ! n1,n2: connected(n1,n2)<-edge(n1,n2).
    ! n1,n2: connected(n1,n2)<-edge(n2,n1).}
\end{lstlisting}

Its translation is:
\begin{lstlisting}[language=IDP, label={lst:2}, caption=Translation of definition]
    ! n1, n2: connected(n1,n2) 
        <=> edge(n1,n1) | edge(n2,n1).
\end{lstlisting}

However, this translation is not ideal for the explanation inference: its result will refer to the equivalence as a whole, and not to the individual rule that applies. 
To address this issue, an alternative translation of the definition is used for the relevance inference:  each rule is translated into one formula, and a completion formula is added, as follows.
\begin{lstlisting}[language=IDP, label={lst:explain}, caption=Translation for explanation]
  ! n1, n2: connected(n1,n2)<=edge(n1,n2).
  ! n1, n2: connected(n1,n2)<=edge(n2,n1).
  ! n1, n2: connected(n1,n2)
      => edge(n1,n1) | edge(n2,n1).
\end{lstlisting}
(This alternative translation cannot be used in the other inferences without adversely impacting response time.)

\textbf{Inductive definitions} of predicates are reduced to formulae acceptable by Z3, as explained by~\citeauthor{pelov2005reducing}~(\citeyear{pelov2005reducing}).
A \emph{level mapping} function symbol is added to the vocabulary for every inductively defined predicate symbol.
This function  associates a real number to each tuple of arguments in the domain of the defined predicate.
This number is related to the stage at which the inductive process determines the interpretation of the predicate for that tuple.
Inequalities on level mappings are added to the theory to ensure that the value of a defined atom is derived only when it is ``safe'' to do so, i.e., when it is known that the condition that triggers the rule will continue to hold~\cite{DBLP:journals/ai/0001VD18}.
Variations of these inequalities are used to support the completion, well-founded, co-inductive and Kripke-Kleene semantics of inductive definitions.
The well-founded semantics is used by default.

\textbf{The determination of relevant atoms} is important for interactive applications, such as the Interactive Consultant (described in the next Section) because it helps reduce the amount of information that the user has to enter. 

An atom is relevant when it is an essential part of at least one ``solution''.
Formally, a {\em solution} of T for S is a partial structure more precise than S such that all of its (total) expansions are models of T.
A {\em minimal solution} of T for S is a solution of T that is not strictly more precise than another solution of T.
An atom is {\em relevant} for theory T and structure S if it has a definite value in a minimal solution of T for S.  

One method to determine relevant atoms is thus to collect the atoms in all minimal solutions of T for S. 
This is, however, not computationally efficient.

Instead, IDP-Z3 uses a 3-step approach:
1) it determines all the consequences of T and S, using the propagation inference;
2) it replaces all occurrences of atoms in the grounded version of T\footnote{The grounded version is obtained by expanding quantifications and aggregates over their (finite) domain.} by their values, if known in S or by propagation, and 
3) it simplifies the resulting theory using the laws of logic.
The atoms that do not occur in the resulting formula are necessarily irrelevant.
Indeed, a solution interpreting any of them cannot be minimal. 

Some irrelevant atoms may not be eliminated by the simplification process.
For example, our implementation fails to simplify $p() \lor q() \lor \lnot p$ to $q()$, and thus incorrectly considers $p()$ relevant.
Hence, the method yields an over-approximation of relevant atoms.

With this method, some atoms involving defined symbols are considered relevant, even though they do not have a definite value in a minimal solution.
We thus refined the method as follows.
%
Recall that an \fodot theory has 3 types of elements: formulas, (inductive) definitions and enumerations.
After simplification (step 3), we obtain simplified formulas and simplified definitions.
We consider a ground atom relevant if, and only if, (i) it is of interest to the user, or (ii) it occurs in a simplified formula, or (iii) it occurs in the simplified definition of a ground atom previously considered relevant.
This method has many similarities with the one implemented in IDP3~\cite{ijcai/JansenBDJD16}.

For example, consider a theory that has some formulas and the only definition of Listing~\ref{lst:1}.  
\lstinline{edge(A,B)} is relevant only if it occurs in a simplified formula of the theory, or if it occurs in the simplified definition of \lstinline{connected(A,B)}, and \lstinline{connected(A,B)} is relevant.

\section{Interactive Consultant}\label{sec:IC}

\begin{figure*}
\begin{center}
    \includegraphics[width=0.95\textwidth]{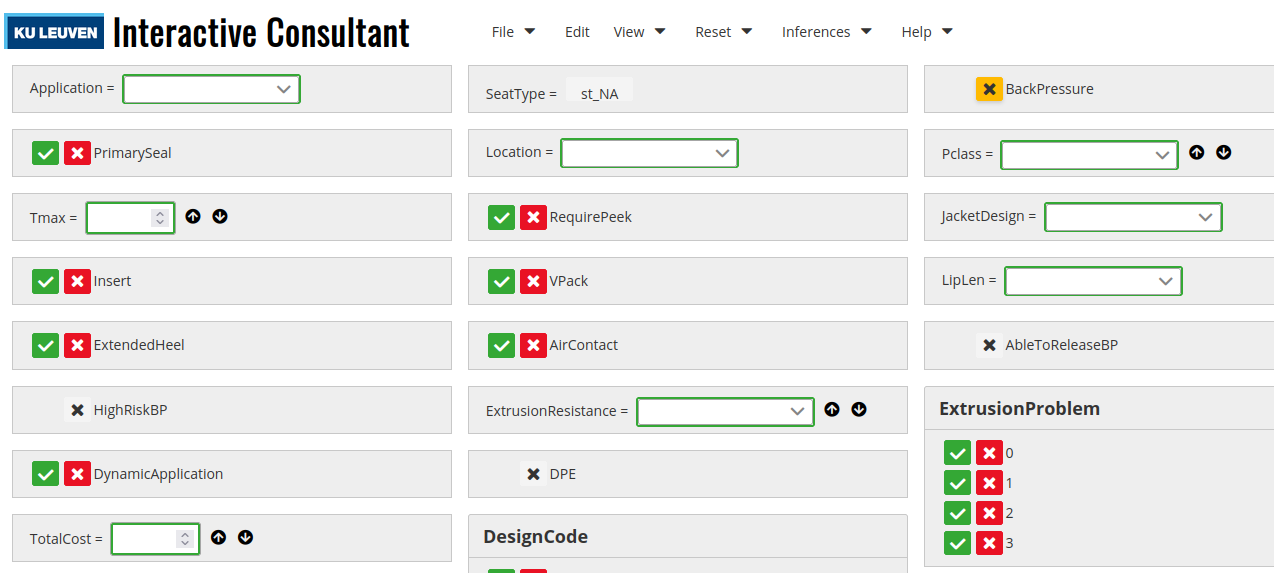}
    \caption{The Interactive Consultant asks relevant questions to its user, from which it draws conclusions that it can explain.}
    \label{fig:ic}
\end{center}
\end{figure*}

A configuration problem can be understood as the search for a state of affairs that extends a fixed partial state of affairs, the \emph{environment}, while satisfying certain conditions that are formally specified.
In a product configuration problem, the environment is the set of customer requirements, and the conditions are the feasibility constraints that the final product design must satisfy.

The Knowledge Representation methodology to solve such problems is to design a vocabulary to represent the relevant objects and concepts in the states of affairs, and to express the conditions as a logic theory.
The problems then become a model expansion problem, with the theory and the structure describing the environment as input.

When the environment is fully known, model expansion can be automated. 
However, the environment is rarely fully known at the start of the search.
In that case, an interactive system equipped with the knowledge of the logic theory can guide its user in the search for a product configuration, by, e.g., asking relevant questions, or guaranteeing that only feasible configurations are considered.

AutoConfig~\cite{Dasseville2016} is such an interactive system based on the reasoning capabilities of IDP3.
Its successor, the Interactive Consultant~\cite{Carbonnelle2019}, is based on IDP-Z3.
IDP-Z3 improves on IDP3 by its handling of \emph{environmental variables}, and by its use of \emph{incremental propagation} for better performance.

Because decisions are made by a user in an environment, it is indeed important to separate the vocabulary describing the environment from the one describing the decisions and their consequences: while the user has control over his decisions, he does not have control over the environment.  
The inferences described in the previous section have been adapted to accommodate this split vocabulary~\cite{carbonnelle2020interactive}.
In particular, the display is updated to highlight the values of environmental symbols determined by the system by propagation: they cannot be assumed to correctly describe the environment; instead, the user must verify them by observing the environment. 
As a result, and unlike its Auto\-Config predecessor, the IC warns the user when a tentative decision he has entered has prerequisites in his environment, which must be verified.
This mechanism provides for a safe exploration of the decision space by the user.

The Interactive Consultant (aka the IC) is generic in the sense that it can be reconfigured by simply changing the \fodot knowledge base:
the user interface is automatically re-generated.
This helps reduce the cost of developing applications significantly~\cite{Deryck2019}.
It is available online\footnote{\url{https://interactive-consultant.idp-z3.be/}}.

The IC displays entry fields for each symbol in the vocabulary, and allows the user to enter data in any order: he can start entering data that describes his goals, his environment, or tentative decisions he would like to make.  
The \textbf{possible values} inference is used to populate the dropdown list.
The user entry is stored in a data theory that is combined with the knowledge base for reasoning.

The following reasoning tasks are performed by IDP-Z3 after any data entry or data retraction (we denote by $T$ the theory in the knowledge base, and by $S$ the data entered by the user at that time):
\begin{itemize}
    \item \textbf{model expansion} of $(T, S)$ to check that the tentative decisions can lead to a model.
    \item \textbf{model propagation} of $(T, S)$ to determine the consequences of the user's entry.  
        The display of the IC is updated to show the propagated values; these values can only be changed after retracting a data entry that justifies them. 
    \item determination of the \textbf{relevant atoms} for $(T, S)$.
        The display is updated to highlight the relevant atoms.
\end{itemize}

If the user is unsure why a literal $L$ is a consequence of its entry and of the theory, he can ask for an \textbf{explanation}.
After having input all values that he deems necessary, the user can ask the IC to fill the remaining entry fields (\textbf{model expansion}) or to show \textit{the optimal} decision that matches the data entered so far, using \textbf{optimization}.

The response time of the system after the user asserts or retracts a fact mostly depends on the speed of the propagation reasoning task.
Propagation is often performed by iterative satisfiability testing~\cite{janota2015algorithms}, i.e., by checking every ground atom to see if it is a consequence of the theory and user input.
We improve the speed of propagation in the IC by using incremental propagation, i.e., by reducing the number of ground atoms to be checked by this process, as follows:

\begin{itemize}
    \item When new facts are asserted by the user, previously propagated atoms do not need to be checked again: indeed, they will remain consequences of the theory and user input;
    \item When facts are retracted by the user, atoms that were not consequences of the theory and user input will remain so, and do not need to be checked again.
\end{itemize}

\section{Evaluation}\label{sec:eval}

In this section, we evaluate IDP-Z3 by demonstrating that it enables experts to build support tools for engineering applications.
We describe the system's practical applicability by way of a demanding use case and highlight the relevant aspects of IDP-Z3 that were required.
We also briefly touch on its computational efficiency.

\subsection{Component Designer Use Case}\label{sec:component}

\begin{table*}[t]
    \centering
    \begin{minipage}{\linewidth}
        \centering
        \begin{tabular}{l  c c c | c c}
            \toprule
            \multicolumn{1}{c }{Use case} & \# symb     & model size & \# sentences & load time (sec) & prop. time (sec) \\ 
            \midrule
            Component designer         & 92  & 227286 & 42  & 32.9  &   2.8    \\ 
            
            Adhesive selection        & 136 & 2061  & 154   & 8.3  & 2.5    \\
            
            
            Notary          & 31 & 31 & 15 & 0.3 & 0.1 \\
            \bottomrule
        \end{tabular}
        \caption{Propagation time is below 3 seconds, and load time is acceptable. }
        \label{tab:usecases}
        \vspace{2em}
    \end{minipage}
\end{table*}

\citeauthor{Aerts2022} (\citeyear{Aerts2022}) describe the creation of an IDP-based knowledge base system for the design of machine components, implemented in partnership with a multinational company.
This company employs engineers worldwide to conceive ``design-to-order'' components.
Before the collaboration, each engineer followed their own \textit{ad hoc} design process, mostly based on their own experience and preferences.
This approach has multiple downsides: (a)~designing a component consumes a large amount of time, (b)~engineers may choose suboptimal designs, and (c)~if a senior engineer leaves, a great deal of knowledge is lost by the company.

To overcome these issues, the company wanted to boost their engineers with a decision support tool.
For this tool to be useful and accepted by the engineers, it had to meet the following requirements:
\begin{itemize}[labelindent=\parindent,leftmargin=*]
    \item[R1] Reason on in-house knowledge
    \item[R2] Support for (non-linear) arithmetic
    \item[R3] Interactive and easy to use
    \item[R4] Explainable
    \item[R5] Sufficient computational efficiency
\end{itemize}

Originally, the IDP3 system was used to tackle this use case.
However, as we discuss later, the limitations of IDP3 led us to later switch to IDP-Z3.

To satisfy R1, the design knowledge was formalized through a series of knowledge extraction workshops. 
Each workshop spanned a few days, during which both knowledge engineers and design engineers were present.
In total, the finalized KB contains 10 parameters describing 60 different materials (such as a maximum temperature of steel) and 27 parameters for 31 components (such as torque and maximum pressure).

An important aspect of this use case is the extensive use of floating point calculations, non-linear functions over continuous variables and divisions.
For example, the knowledge base contains formulas of the following forms:
\begin{lstlisting}[language=IDP, label={lst:SG_example}, caption=Component use case example formula]
E() = 1117.36 + -9.33 * T() +  0.0374 * T() * T() + -0.0002 * T() * T() * T() + ...
L() = (O() - I() - L(Spring())/2)*0.95
\end{lstlisting}
R2 is met by IDP-Z3 due to its reliance on an SMT solver internally instead of a SAT solver.
In the original IDP3 application, these difficult formulas could only be approximated using an unintuitive work-around, which meant that the results were unreliable.


R3 and R4 are important requirements for the engineers who use the tool.
Indeed, they want a tool capable of supporting them interactively, i.e., \textit{live} during the design process so that it can actively guide them towards correct designs.
Here, IDP-Z3's \textbf{propagation}, \textbf{relevancy} and \textbf{optimization} inferences offer the required functionalities for this interactivity.
Moreover, if during the design process an engineer is unsure about a system-derived value, the system can explain why it was derived using the \textbf{explanation} inference.
By exposing these functionalities through the Interactive Consultant interface, as discussed in Section~\ref{sec:IC}, the interaction with IDP-Z3 becomes accessible to the design engineers, thereby satisfying R3 and R4.
As will be shown in Section~\ref{ss:perf}, the described approach also meets requirement R5.

Overall, the company and its engineers are very positive about the tool.
Besides a reported daily time-save of up to 30 minutes for each engineer, they report other benefits to its usage.
First and foremost, it leads to more ``first-time right'' designs, which lowers production time and cost.
Second, for new engineers, the tool serves as an excellent learning tool, allowing them to indirectly learn from the knowledge of the more experienced engineers.
For more experienced engineers, on the other hand, the tool is used to challenge their assumptions: when in doubt, they can swiftly verify if their initial ideas are correct.
Lastly, with their knowledge captured in a KB, engineers leaving will not result in loss of knowledge for the company.

A similar engineering-focused use case is described in~\cite{Jordens2021,Vandevelde2022}.
Here, a company that employs adhesive experts wants a tool to support them in selecting correct adhesives for gluing operations.
The requirements for this tool are identical to the ones of the component design use case.
The KB was formalized during multiple knowledge workshops through collaboration between adhesive experts and knowledge engineers.
Through the Interactive Consultant interface, the adhesive experts interact with the knowledge.
They especially appreciate being able to explore the problem domain interactively: each time they enter a design requirement, they observe the number of suitable adhesives going down, showing them that they are converging towards a solution.
Selecting adhesives is now also faster, with a reduction in selection time from three hours to five minutes.

Besides the two engineering use-cases outlined in this section, IDP-Z3 has also been successfully used in other domains, such as a legal tool to support notaries~\cite{Deryck2019} and a financial tool for collateral management~\cite{Deryck2021}.
This latter tool currently runs in production at a leading global financial agency~\cite{securitiesfinancetimes_2022}, where it already dramatically cut down the time for one operation from multiple months to mere seconds~\cite{DeryckThesis}.

\subsection{Computational efficiency}\label{ss:perf}


\begin{table*}[t]
    \centering
    \begin{minipage}{\columnwidth}
        \centering
        \begin{tabular}{l  c c c c}
            \toprule
            \multicolumn{1}{c }{Language} & \fodot     & SMT & ASP & mzn\\ 
            \midrule
            \textbf{Classical syntax}          & \checkmark & \checkmark & & \checkmark          \\ 
            \textbf{Types, quantification}        & \checkmark & \checkmark &      & \checkmark      \\
            Uninterpreted types            &          & \checkmark &            \\
            (Inductive) \textbf{definitions}         & \checkmark &           & \checkmark  \\
            Disjunction in the head           &            &           & \checkmark  \\
            \textbf{Linear arithmetic}            & \checkmark & \checkmark & \checkmark & \checkmark \\
            \textbf{Non-linear arithmetic}              & \checkmark & \checkmark &   -$^*$    & -$^*$      \\
            \textbf{Aggregates}                    & \checkmark & -$^*$     & \checkmark & \checkmark \\ 
            Vector, Array                &          & \checkmark &     & \checkmark         \\
            Concept as a type            & \checkmark &           &            \\
            \bottomrule
        \end{tabular}
        \caption{Comparison of model-based languages. Features in bold are requirements for the Component Use Case.\\
            $^*$Not part of the standard, but provided by some solvers.}
        \vspace{2em}
        \label{tab:languages}
    \end{minipage}
    \begin{minipage}{\columnwidth}
        \begin{tabular}{l  c c c c}
        \toprule
        \multicolumn{1}{c }{Reasoning} & IDP-Z3     & SMT & ASP & mzn\\
        \midrule
        \textbf{Model checking}          & \checkmark & \checkmark &  \checkmark   & \checkmark       \\ 
        \textbf{Model expansion}        & \checkmark & \checkmark &  \checkmark  & \checkmark        \\
        \textbf{Possible values}           &  \checkmark        &  &   &          \\
        \textbf{Propagation}           &  \checkmark        & -$^{*}$ &            \\
        \textbf{Explanation (unsat)}           & \checkmark & \checkmark &  & \checkmark \\ 
        Step-wise explanation   & -$^{*}$ &  &   \\
        \textbf{Optimization}         & \checkmark &  -$^{*}$         & \checkmark & \checkmark \\
        \textbf{Relevance}           & \checkmark           &           &   \\
        \bottomrule
        \end{tabular}
        \caption{Comparison of model-based reasoning systems. Features in bold are requirements for the Component Use Case.\\
            $^*$Provided by some solvers.}
        \label{tab:reasoning}
    \end{minipage}

    \begin{minipage}{0.55\linewidth}
        \centering
        \begin{tabular}{cccccc}
        \toprule
             System & R1 & R2 & R3 & R4 & R5   \\
             \midrule 
             SMT (Z3, CVC5, $\ldots$) & ++ & ++  & & ++ & ++ \\
             ASP (Clingo, DLV, $\ldots$) & + & + & &  & ++ \\
             MiniZinc (Gecode, ...) & ++ & + & & ++ &  \\
             IDP3 & ++ & - & ++ & ++ & ++ \\
             \midrule
             IDP-Z3 &  ++ & ++ & ++ & ++ & + \\
             \bottomrule 
        \end{tabular}
        \caption{Comparison between model-based systems on the requirements of the Component Use Case presented in Section~\ref{sec:component}}
        \label{tab:reqs}
    \end{minipage}
    
\end{table*}

We now discuss the computational efficiency of IDP-Z3 in our case studies.
We believe that these results are representative for other interactive applications based on IDP-Z3 and the Interactive Consultant.

Table~\ref{tab:usecases} shows various metrics and speed indicators for each use case.
The column headings are:
\begin{itemize}
    \item \# symb: the number of symbols in the vocabulary;
    \item model size: the size of a model, i.e., the sum of the cardinality of the domain of each predicate and function symbol (they all have a finite domain);
    \item \# sentences: the number of formulas and definitional rules in the theory;
    \item load time: the number of seconds needed to load the knowledge base in the Interactive Consultant;
    \item prop. time: the number of seconds needed to propagate the assertion or retraction of a fact.
\end{itemize}

The load and response times are measured on an Intel® Core™ i7-8850H CPU @ 2.60GHz × 12 machine, with 16 GB of memory, using Ubuntu 20.04.03, CPython 3.9 and IDP-Z3 0.9.2.

The table shows that the load time can exceed 10 seconds for KB with large models.
This is not ideal, but still acceptable in the use case presented.
The delay is due to the transformation of the \fodot KB into the equivalent grounded FO formula that is submitted to Z3: the transformation is performed at initial load, by code written in Python, a language not known for its speed.  
Load time could be improved by rewriting the transformation in a faster language, or by loading the pre-computed result of the transformation directly from storage.

The response time of the Interactive Consultant, on the other hand, is below 3 seconds, an adequate delay for interactive applications.
This is achieved by retaining the internal state of IDP-Z3 between interactions. 
Our experience indicates that such retention is important in interactive applications based on IDP-Z3.
Response time could be further improved by parallelizing propagation, i.e., by running the iterative satisfiability testing method over multiple instances of Z3.


\begin{table*}
\end{table*}

\section{Related Work}\label{sec:related}

We now discuss whether other reasoning engines could be used to build an interactive consultant for the Component Use Case and other similar cases.
We consider reasoning engines for three other model-based languages: SMT-LIB-2~\cite{SMT}, ASP-Core-2~\cite{ASP} and MiniZinc~\cite{DBLP:conf/cp/NethercoteSBBDT07}.



Table~\ref{tab:languages} compares \fodot to the other languages.
Table~\ref{tab:reasoning} compares IDP-Z3 to reasoning engines for the other languages, based on the required functionality of an interactive configurator in~\citeauthor{VanHertum2017} (\citeyear{VanHertum2017}).
The features in the comparison tables are described in Sections~\ref{sec:fodot} and~\ref{IDP-Z3}.
Table~\ref{tab:reqs} shows that IDP-Z3, unlike other systems, meets all requirements of our Use Case.

SMT solvers are closest to what the use case needs: in fact, this is the reason we used the Z3 solver as back-end.
However, SMT solvers do not provide the relevance inference: yet, interactive configurators require the capability to ask relevant questions (for R3) because it reduces the effort of their users. 
As described in Section~\ref{IDP-Z3}, the relevance inference must make a distinction between constraints and (constructive) definitions. 
Since \fodot makes such a distinction, it is superior to SMT-LIB-2.
\fodot also supports aggregates, which are useful in configuration problems to, e.g., compute the total cost of a configuration.
On the other hand, SMT-LIB-2 supports uninterpreted types, which is useful in so-called generative configuration problems, i.e., problems in which the number of some components is not known.

ASP-Core-2 supports arithmetic, but programs must be ``safe'' and have a finite number of answer sets~\cite{ASP}.
Thus, ASP solvers do not allow reasoning in case where the value of some physical parameters is not fixed by the constraints.
Yet, when using an interactive application, the user progressively enters information: hence, many parameters initially do not have a fixed value, and some parameters may remain unknown.
Still, the interactive application must determine the consequences of the user's entry (by the propagation inference).
Thus, ASP solvers do not meet requirement (R3). 
Note, however, that the safety and finiteness constraints are relaxed in Constraint ASP (CASP) and ASP Modulo Theories (ASPMT) solvers, which supports linear arithmetic (R2).

Additionally, ASP solvers do not have support for the explanation (unsat\_core) nor the relevance inference, which are important in interactive applications (R3, R4).

\ignore{Furthermore, they do not support types in the sense that predicate symbols are defined over the whole domain.
Types are used to determine the possible values of a term when generating the user interface; their absence makes this generation more difficult (R3).}

MiniZinc is another expressive formalism, but it does not support inductive definitions (which is not necessary in our use case, but is necessary in network configuration problems where the reachability relation must be computed).
Some MiniZinc solvers support non-linear arithmetic (R2).

The FindMUS command of MiniZinc can be used to develop the explanation inference.  
For the relevance inference, the MiniZinc formalism is weaker than the \fodot formalism as it does not differentiate constraints from definitions, as explained for SMT above.

Unlike SMT and ASP solvers, MiniZinc solvers do not support incremental solving (called multi-shot ASP in ~\citeauthor{gebser2019multi}): when a constraint is added to a theory, a new search for a solution is started, ignoring solutions found for the original theory.
This has an adverse effect on the performance of the Relevant Values and the Propagation inferences (R5).

As stated before, IDP3 is very similar to IDP-Z3, but lacks support for linear arithmetic (R2).

\ignore{
    None of the languages or systems are complete.  
    Hence, they are under further development to bring more expressivity to the languages, and more reasoning capability to the reasoning engine.
    
    Because they share many concepts, some researchers have investigated the possibility to transform a KB in one language into a KB in another, e.g., to improve performance:
    \begin{itemize}
        \item IDP-Z3 itself transforms \fodot KBs into SMT-LIB-2 KBs;
        \item FOLASP transforms \fodot KBs into ASP-Core-2, allowing performance comparisons~\cite{VanDessel2021}; the semantics correspondence between FO(ID) and ASP is explored in~\citeauthor{DBLP:conf/iclp/DeneckerLTV12} (\citeyear{DBLP:conf/iclp/DeneckerLTV12});
        \item and several ASP-Core-2 solvers are based on SMT solvers (e.g.,~\citeauthor{DBLP:conf/iclp/ShenL18}).
    \end{itemize}
    
    We expect these investigations to continue.

    Below are some additional notes.
    
    \paragraph{Classical notation}
    Unlike \fodot and SMT-LIB-2, ASP-core-2 has been strongly influenced by logic programming.  
    As a consequence, its syntax is significantly different from first-order logic.

    The below is not correct: "{notvote; a18}." has the correct models in ASP.  try it here: https://potassco.org/clingo/run/ 

    The lack of classical logical connectives in ASP-Core-2 makes the language less expressive than \fodot and SMT.
    For example, the non-mandatory voting law of Listing~\ref{lst:non-prescriptive} cannot be represented in ASP (without adding more symbols).
    Indeed, the following ASP program has only two models (\lstinline|{notvote}| and \lstinline|{a18}|), and is missing (\lstinline|{notvote, a18}|).
    \begin{lstlisting}[language=Prolog, label={lst:ASP}, caption={Non-mandatory voting, in ASP}]
        notvote | a18.
    \end{lstlisting}
}
\ignore{
    \paragraph{Elaboration tolerance}
    Elaboration tolerance~\cite{mccarthy1998elaboration} is an important property of Knowledge Representation languages.
    In a language that is elaboration tolerant, it is easy to change a formulation to take into account new circumstances, and in particular new knowledge (\emph{additive elaboration}).
    We already explained in Section~\ref{sec:fodot} that ``Concept as a type'' makes \fodot more elaboration tolerant by eliminating the need for reification.
    We contend that ASP is less elaboration tolerant than \fodot and SMT-LIB because of its minimality condition: adding new knowledge can lead to unexpected results.
    \todo{add an example, or drop the last sentence}
}

\ignore{
    \paragraph{\fodot vs. IDP}
    Note that Table~\ref{tab:languages} evaluates the \emph{declarative} logic language \fodot, while Table~\ref{tab:reasoning} evaluates the \emph{imperative} reasoning language IDP. 
    SMT-LIB-2 and ASP-core-2 do not make that distinction: they combine the declarative and imperative languages.
    (In fact, the IDP acronym comes from ``Imperative + Declarative Programming'')
}

\ignore{
    Below are some additional notes.
    
    \paragraph{Classical notation}
    Unlike \fodot and SMT-LIB-2, ASP-core-2 has been strongly influenced by logic programming.  
    As a consequence, its syntax is significantly different from first-order logic.

    The below is not correct: "{notvote; a18}." has the correct models in ASP.  try it here: https://potassco.org/clingo/run/ 

    The lack of classical logical connectives in ASP-Core-2 makes the language less expressive than \fodot and SMT.
    For example, the non-mandatory voting law of Listing~\ref{lst:non-prescriptive} cannot be represented in ASP (without adding more symbols).
    Indeed, the following ASP program has only two models (\lstinline|{notvote}| and \lstinline|{a18}|), and is missing (\lstinline|{notvote, a18}|).
    \begin{lstlisting}[language=Prolog, label={lst:ASP}, caption={Non-mandatory voting, in ASP}]
        notvote | a18.
    \end{lstlisting}
}
\ignore{
    \paragraph{Elaboration tolerance}
    Elaboration tolerance~\cite{mccarthy1998elaboration} is an important property of Knowledge Representation languages.
    In a language that is elaboration tolerant, it is easy to change a formulation to take into account new circumstances, and in particular new knowledge (\emph{additive elaboration}).
    We already explained in Section~\ref{sec:fodot} that ``Concept as a type'' makes \fodot more elaboration tolerant by eliminating the need for reification.
    We contend that ASP is less elaboration tolerant than \fodot and SMT-LIB because of its minimality condition: adding new knowledge can lead to unexpected results.
    \todo{add an example, or drop the last sentence}
}

\ignore{
    \paragraph{\fodot vs. IDP}
    Note that Table~\ref{tab:languages} evaluates the \emph{declarative} logic language \fodot, while Table~\ref{tab:reasoning} evaluates the \emph{imperative} reasoning language IDP. 
    SMT-LIB-2 and ASP-core-2 do not make that distinction: they combine the declarative and imperative languages.
    (In fact, the IDP acronym comes from ``Imperative + Declarative Programming'')
}

\section{Conclusions}\label{sec:conclusion}

Our work shows that interactive applications based on IDP-Z3 deliver real value to their users.
They can be developed at low costs (typically 10 days of effort) and still exhibit acceptable performance (response time typically below 3 seconds).  

The following elements have contributed to this success:

\begin{itemize}
    \item Our use cases are knowledge-rich but data-poor, making it possible to use computationally complex forms of reasoning.  IDP-Z3 brings value by having knowledge that the user may not have, and by reasoning faster and more rigorously with it than an expert can.  
    \item IDP-Z3 allows reasoning not only with deterministic rule-based definitions, but also with non-deterministic formulas describing possible worlds.  Our result further justifies the revival of interest in the seminal papers on the integration of rule-based languages and classical logic (\citeauthor{DeneckerCL2000} (\citeyear{DeneckerCL2000}), 20-year Test-of-Time award at ICLP 2020, and \citeauthor{DBLP:conf/iclp/DeneckerPB01} (\citeyear{DBLP:conf/iclp/DeneckerPB01}), 20-year Test-of-Time award at ICLP 2021).
    \item The use of generic reasoning methods (as recommended in the Knowledge Base paradigm) and the automatic generation of the user interface of the Interactive Consultant significantly reduce the development costs of intelligent applications.  
\end{itemize}

The interaction between the user and the Interactive Consultant has many similarities with the conversation in a Turing test.
Here, the interaction is not conducted in a natural language, but the machine shows signs of intelligence that would be tested in a Turing test, such as the capability to ask relevant questions or provide explanations.
This insight justifies further research.

Additional research would expand the expressiveness of the knowledge representation languages and/or improve the performance of reasoning engines.
In particular, support of partial functions is beneficial in many practical applications, and support of uninterpreted types is beneficial for generative configuration problems~\cite{DBLP:journals/expert/FleischanderlFHSS98}.

\ignore {
We have presented IDP-Z3 and its use in four knowledge-intensive, interactive applications, ranging from engineering to finance and law.
    We have compared it to other related systems.
    
    In future work, \fodot could be standardized to facilitate collaboration with other research teams. 
    Work is ongoing to develop IDP-Z3 in these directions:
    \begin{itemize}
        \item support for \emph{partial functions}, i.e., for functions that are undefined for some arguments.  They often occur in practical applications (e.g., the \idp{husband : People -> People} function).
        \item reasoning with uninterpreted types, i.e., types whose extension is not known in advance.
        \item use other back-end solvers.
    \end{itemize}
    
    In the near future, we plan to use IDP-Z3 to teach Knowledge Representation at graduate level.
    
    \marc{
    1) Much efforts on KR and computational logic go to BIG applications and scalability, which leads to strong focus on computational complexity of logics and reasoning. But the applications described here are all SMALL but knowledge intensive, making it feasible to use computationally complex forms of inference. Small means : small in search space, but that does not means easy. Humans users of the system experience great help from a system like this that understands the domain knowledge, and guards over it when solutions are being computed. After all, humans are very poor reasoners. 
    
    2) Part of the success of iIDP=Z3 is the integration of classical logic and rule-based representations. interpreted as definitions, leading to a language offering 2 non-equivalent conditionals. Tools for classical logic and rule-based langauges are typically  separate worlds,  with rule-based systems being bound to very spcific inference tasks making them unsuitable for reasoning with uncertainty, or reasoning backwards from desired conclusions to constraints on input.  In \fotodot, these two represenation paradigms are cleanly integrated. Maybe here it could be said that the original paper on \fodot won the 20 Y Test of time award?  It seems like an opportunity to play trump.
    
    3) The context interactive systems is extremely interesting for research in KRR. Not in the last place, because an intelligent system that understands the domain knowledge, can and should provide so many different functionalities to support  a user.  It is here that the knowledge base paradigm, with its view of reusing a KB using different sorts of inference, serves best of all applications that I have encountered. It also in this sort of application that symbolic AI, with its spectrum of different forms of reasoning on the same knowledge could play a role that cannot soon be played by neural network systems. Furthermore, the   interaction between a human user and an intelligent interactive system   has also something in common with Turings test with the conversation between a human and an artificial mind.  Here, the interaction is not conducted in natural language, but the way the  system  supports the user by propagating,  explaining, checking consistency, generating examples, optimisation, etc.: they are signs of intelligence . 
    
    }}

\section*{Acknowledgements}
This research received funding from the Flemish Government under the ``Onderzoeksprogramma Artificiële Intelligentie (AI) Vlaanderen'' programme.

The authors thank Ingmar Dasseville and Jo Devriendt for their contributions to the development of IDP-Z3.
They also thank the contributors to the many open source projects IDP-Z3 is built on, e.g., Z3 or textx~\cite{textx}.

\bibliographystyle{kr}
\bibliography{biblio,krrlib}

\end{document}